\documentclass{PoS}

\usepackage{wrapfig}
\newcommand{\pom}{I\!\!P}

\newcommand{\xpom}{x_{\pom}}

\newcommand{\rcs}{\sigma_r^{D(3)}}
\newcommand{\rcsfull}{\sigma_r^{D(3)}(Q^2, \beta, \xpom)}
\newcommand{\ud}{\mathrm{d}}

\title{Measurement of inclusive diffractive deep inelastic scattering using VFPS at H1}

\ShortTitle{Measurement of inclusive diffractive deep inelastic scattering using VFPS at H1}

\author{\speaker{Tom\'a\v{s} Hreus}%
         \thanks{on behalf of H1 Collaboration}\\
        Universit\'e Libre de Bruxelles\\
        E-mail: \email{hreus@mail.desy.de}}

\abstract{Performances of the Very Forward Proton Spectrometer (VFPS) of the H1 detector at HERA using data
collected during the 2006/2007 running period are discussed, including the description of acceptance,
reconstruction of proton energy loss and estimation of the amount of the beam-gas background.
The first physics result obtained with the VFPS detector -
a preliminary measurement of the semi-inclusive reduced cross section,
$\rcsfull$ - is presented
for the diffractive deep inelastic scattering process $ep \to e X p$ with the leading final state
proton measured by the VFPS. 
Results of this measurement are found to be in agreement with other $\rcs$ measurements by the H1 Collaboration
and with a theoretical prediction based on a NLO DGLAP QCD fit.
They provide a higher precision and therefore allow a deeper inside to the nature of diffraction.}

\FullConference{XVIII International Workshop on Deep-Inelastic Scattering and Related Subjects\\
                 April 19 -23, 2010\\
                 Convitto della Calza, Firenze, Italy}

\begin{document}

\section{Introduction}
\begin{wrapfigure}{r}{0.3\columnwidth}
\centerline{\includegraphics[width=0.3\columnwidth,clip=]{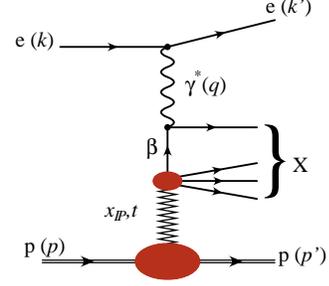}}
\caption{Diffraction diagram}
\label{Fig:DiffDiag}
\end{wrapfigure}
Diffractive processes have been studied extensively in deep-inelastic electron-proton scattering (DIS) at
the HERA collider (see e.g.~\cite{Aktas:2006hy,Aktas:2006hx}) and
are characterised by a colour singlet exchange between the proton and electron.
They result in two distinct hadronic systems in the final state separated by a large rapidity gap: the dissociation system of the photon
and the scattered proton (Fig.~\ref{Fig:DiffDiag}).
The high energy of HERA beams offers the opportunity to try to understand the diffractive exchange taking place here
in terms of a fundamental theory, QCD, in the perturbative regime, where the hard scale is supplied by 
the large photon virtuality $Q^2 = -q^2$.
Along with kinematic variables defined in inclusive DIS, further variables specific to diffraction are 
the momentum fraction of proton carried by the exchanged colour singlet, $\xpom = q\cdot (p - p') / q\cdot p$, 
the momentum fraction of the colour singlet carried by the struck quark, $\beta = -q^2 / q \cdot (p - p')$
and the momentum transfer squared at the proton vertex, $t = (p - p')^2$.
Several methods exist to experimentally measure diffraction. In the large rapidity gap method, events are selected on the basis
of a large gap between photon dissociation system $X$ and the direction of scattered proton. 
However, the remaining fraction of the proton dissociative events in the sample translates into
a dominant systematic uncertainty \cite{Aktas:2006hy}.
On the other hand, one of the advantages of directly measuring the scattered proton is a rejection of the proton dissociative events.
For that reason, a dedicated proton spectrometer with a large acceptance, the VFPS detector, was installed at H1 in 2004.
H1 experiment contains two such proton spectrometers: VFPS and FPS. The FPS detector however, has a limited acceptance
($\sim 5\%$ in the diffractive kinematic region).

\section{VFPS Performance}
\begin{wrapfigure}{r}{0.5\columnwidth}
\centerline{\includegraphics[width=0.45\columnwidth,clip=]{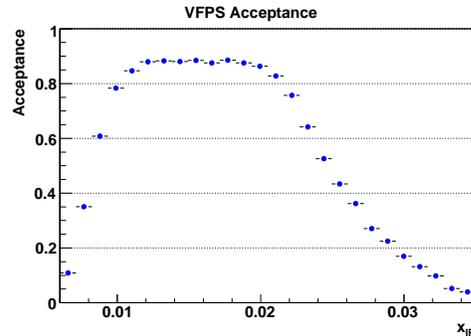}}
\caption{The VFPS acceptance in $\xpom$.}
\label{Fig:Acc}
\end{wrapfigure}
VFPS consists of two retractable ``Roman Pot'' stations 
installed in the 6.2 m long drift section of the proton beam pipe at 218 and 222 m from the interaction point
(in direction of outgoing protons), i.e. in the bending region of the HERA beam.
The Roman Pot principle is based on the construction of an insert into the beam pipe which allows tracking detectors 
to be moved close to the beam, in this case in the horizontal fashion.
Each station is equipped with 2 scintillating fibre detectors and
each such detector measures two track coordinates using two planes of fibres. 
Scintillating fibres are sandwiched between four trigger tiles on both sides, which
deliver a fast level 1 signal to H1.
A complete information on the VFPS construction and installation can be found in \cite{vfps_proposal}.
Acceptance of the VFPS depends on the proton beam optic. 
At 220m from the interaction point, diffracted protons which have lost approximately $1\%$ of their energy
can leave the beam envelope and therefore be detected.
Acceptance in $\xpom$ and $t$
is highest between $0.01 \lesssim \xpom \lesssim 0.025$ and $|t| \lesssim 0.25$ GeV$^2$,
down to the lowest kinematically accessible $t$ values (Fig.~\ref{Fig:Acc}). 
The acceptance further depends on the horizontal distance of Roman Pots from the beam center during run, which affects
the ability of detectors to measure protons with smaller $\xpom$. 
To increase the acceptance in large $\xpom$ region, local beam offsets were applied during 2006-2007 running period, when
the proton beam was moved horizontally and vertically off-center by several millimeters and thus allowed protons with larger energy losses
(deviated more by the bending magnets) to be detected in VFPS. 
The absolute proton beam position was measured by calibrated beam monitors at 220m with precision of $200\mu$m.
From the measured impact points by the scintillating fibers and tiles in the two Roman pot stations, the position
and angle of the scattered proton are determined independently for each station.
VFPS track reconstruction efficiency was measured to be $96\% \pm 4\%$ for a requirement that a track be reconstructed in either of the two VFPS stations.
The $\xpom$ variable is reconstructed assuming the linear approximation and $t = 0$, 
as a distance $\Delta x$ in a horizontal direction between
the beam center and the VFPS impact: $\xpom = A \cdot \Delta x + B$, where $A$ and $B$ are calibration constants.
A cross-check of the $\xpom$ reconstruction has been made on a sample of diffractive $\rho^0$ electroproduction events,
$ep \to e \rho^0 p$, by a comparison between $\xpom^\rho$ reconstructed from two pion tracks and $\xpom$ measured from proton in VFPS,
showing a $5\%$ absolute calibration. 
The resolution on $\xpom$ is $12\%$, a significantly better compared to reconstruction using
information from the main H1 detector, which is at the level of $20\%$. 

\section{Cross Section Measurement of inclusive diffraction in DIS}
\begin{wrapfigure}{r}{0.5\columnwidth}
\centerline{\includegraphics[width=0.45\columnwidth,clip=]{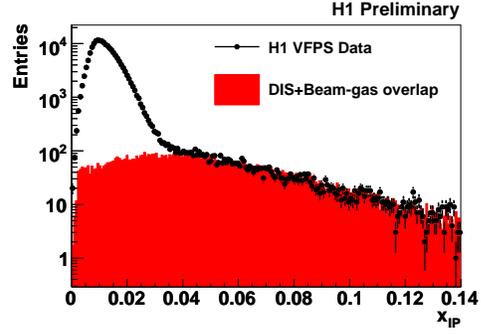}}
\caption{Estimation of the beam-gas+DIS overlap background into the selected diffractive sample. Background is
normalised to data in the region of $\xpom > 0.04$.}
\label{Fig:Bg}
\end{wrapfigure}

\begin{figure}
\begin{center}
\includegraphics[width=0.32\columnwidth]{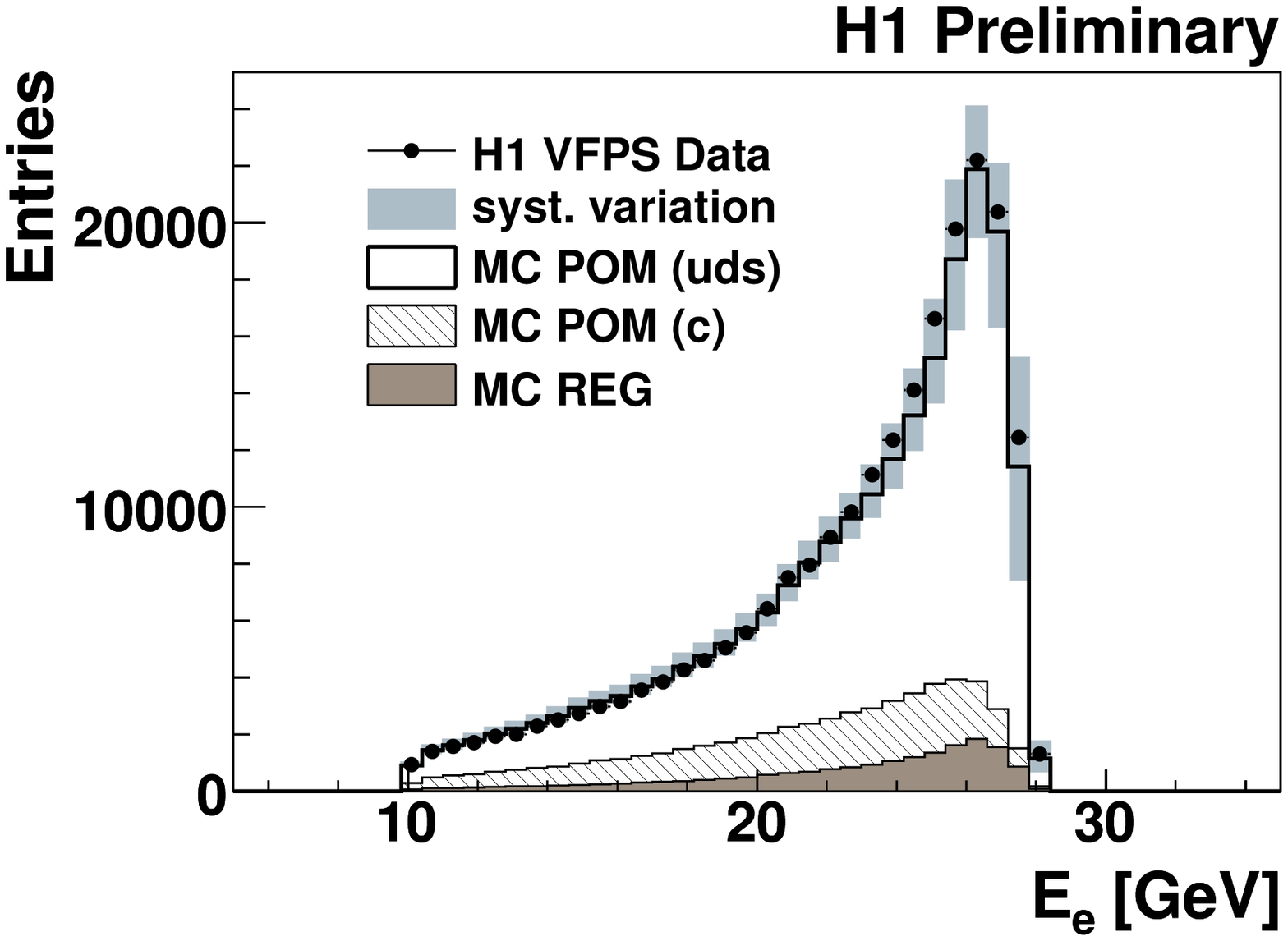}
\includegraphics[width=0.32\columnwidth]{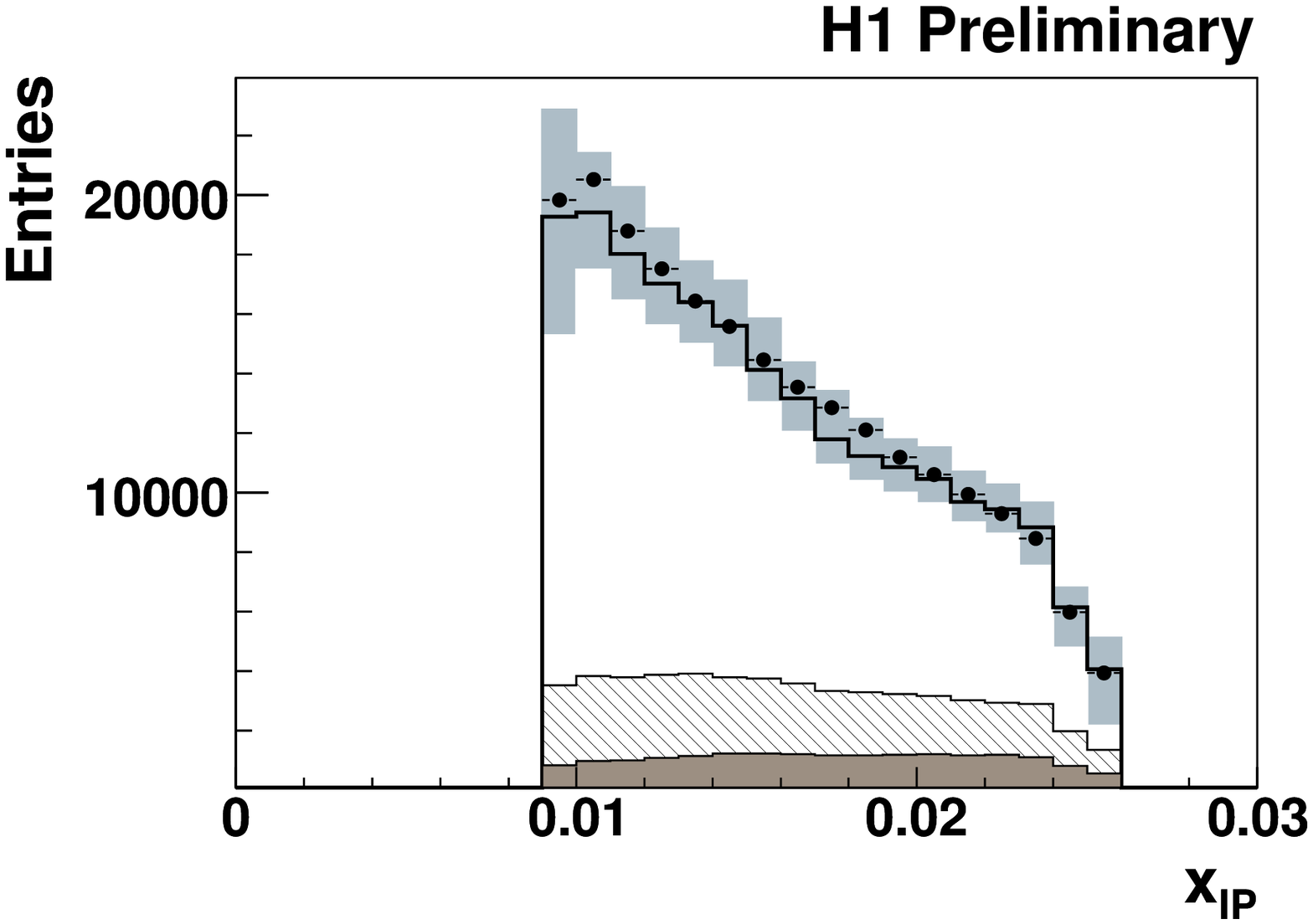}
\includegraphics[width=0.32\columnwidth]{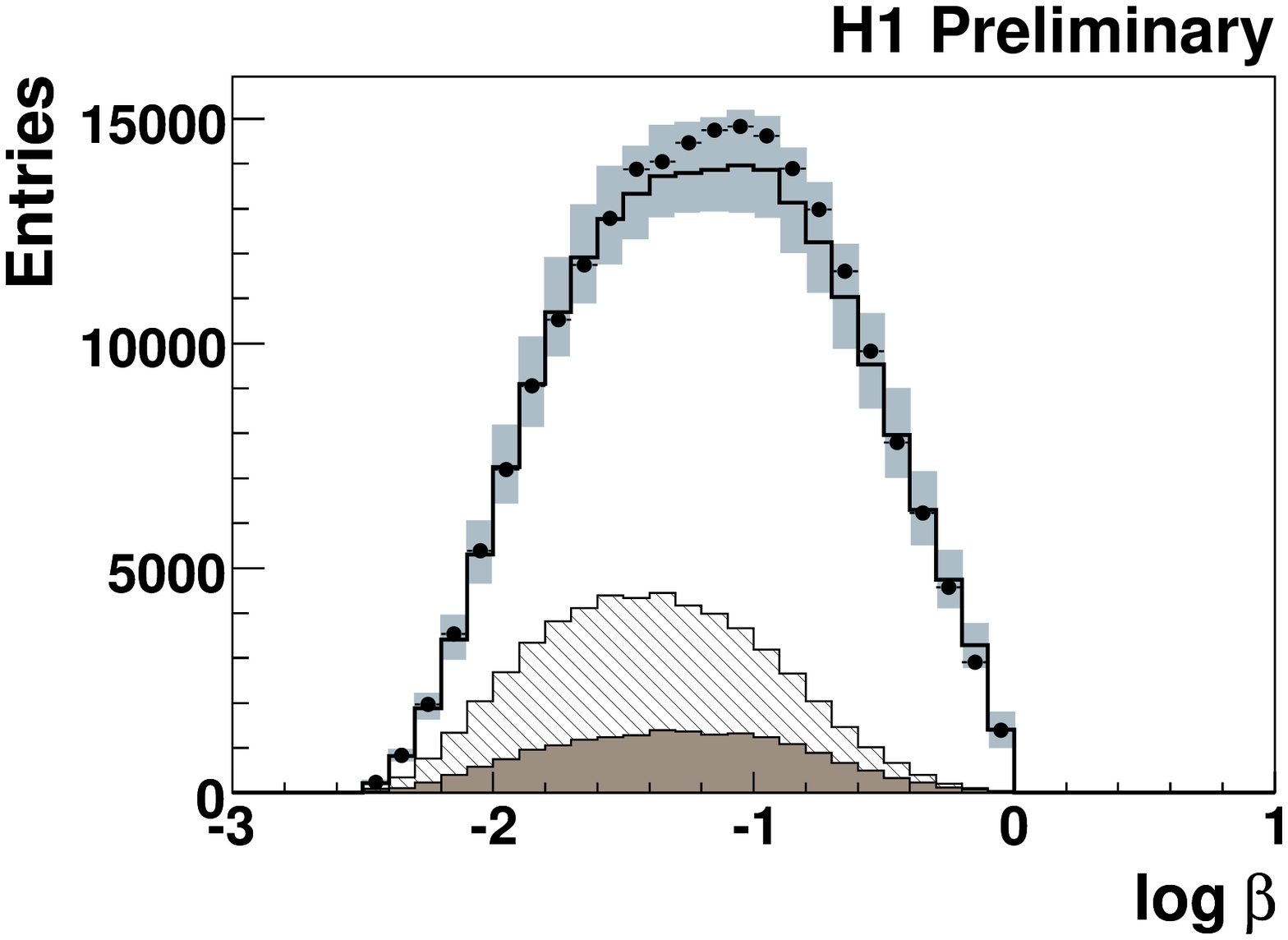}
\caption{Distributions of selected diffractive sample: scattered positron energy $E_e$, $\xpom$ and $\log \beta$.}
\protect\label{Fig:Control}
\end{center}
\end{figure}

The cross section of inclusive diffraction is measured in DIS regime, differentially in $Q^2$, $\beta$ and $\xpom$
in the range $4.5 < Q^2 < 100$ GeV$^2$, $0.008 < \beta < 1$ and $0.009 < \xpom < 0.026$.
Cross section is measured using diffractive events of type $e^+p \to e^+ X p$,
with the scattered proton measured by the VFPS detector. 
Data correspond to an integrated luminosity of $87.4$ pb$^{-1}$ and were collected with the H1 detector during
the 2006-2007 running period, when positrons with energy of $27.6$ GeV collided with protons with energy 
of 920 GeV.
For the event to be selected,
a scattered positron detected as a cluster in the electromagnetic section of the calorimeter is required
to have an energy $E_e > 10$ GeV. Furthermore, a reconstructed event vertex is required 
and the VFPS trigger together with a reconstructed proton track
in either of the two VFPS stations.
The reduced cross section, $\rcs$, is related to the differential cros section measured experimentally by
\begin{equation}
 \nonumber
 \frac{ \ud^3 \sigma^{ep\to eXp} }{\ud Q^2 \ud \beta \ud \xpom}
 =
 \frac{4 \pi \alpha_{em}^2}{\beta Q^4} 
 \cdot \Big( 1 - y + \frac{ y^2 }{ 2 } \Big)
 \cdot \rcsfull ~,
\end{equation}
where $y$ is inelasticity. 
The reconstruction of the inclusive DIS kinematic variables is performed using a mixed method \cite{Aktas:2006hx}, 
where the information is exploited from both the scattered positron and the hadronic final state.
The $\beta$ variable is reconstructed as $\beta = x / \xpom$, where $x$ is the Bjorken scaling variable.
Background from coincidences of DIS events with beam-gas 
protons giving a signal in the VFPS was estimated 
using random combination of DIS events ($ep \to eX$) and beam-gas protons 
($pA \to pA'$) from data events using an independent trigger. 
Fig.~\ref{Fig:Bg} shows the $\xpom$ distribution measured using
information of the main H1 detector, 
for data and beam-gas events normalised 
to data for $\xpom > 0.04$, i.e.~outside 
of the VFPS acceptance. 
The remaining beam-gas contribution at $\xpom < 0.04$ (cut in the analysis),
is estimated to be $2\%$ and is subtracted from the number of selected events in the reduced cross section measurement.
The RAPGAP 3.1 MC generator \cite{Jung:1993gf} is used to simulate the diffractive $ep$ scattering
using predictions based on diffractive parton distribution functions extracted from an NLO DGLAP QCD fit to H1 LRG data \cite{Aktas:2006hy},
H1 2006 DPDF Fit B.
Fig.~\ref{Fig:Control} shows distributions of the scattered positron energy, $\xpom$ as reconstructed from VFPS and $\beta$
compared to MC simulation with total systematic variations.
The MC is normalised to data luminosity after correction for the
different kinematic region of the cross section in H1 LRG data, which includes proton dissociation 
with masses $M_Y < 1.6$ GeV, whereas $M_Y = M_p$ in our case. A global correction of 0.81 \cite{Aktas:2006hx} was used to account for this 
difference. 
An excellent agreement is obtained between data and MC simulation within the systematic uncertainty.
The systematic sources related to VFPS are the
normalisation uncertainty of VFPS trigger efficiency (1\%) and track efficiency (4\%)
and $5\%$ uncertainties of calibration parameters $A$ and $B$ in the $\xpom$ reconstruction, which result in the uncertainty on the reduced
cross section of up to $5\%$ and $8\%$, respectively (uncertainties of $A$ and $B$ were set to conservative values, 
with a room for improvement). 
Model distributions were reweighted by 
$1/\xpom^{\pm 0.1}$, $1/\beta^{\pm 0.05}$ and $e^{\pm t}$, with the resulting uncertainties on the reduced cross section up to $3\%$.
The total VFPS normalisation uncertainty is at the level of $5\%$ 
and the total systematic uncertainty is between $8\%$ and $14\%$. 
The reduced cross section in bin $i$, defined as:
\begin{equation}
 \nonumber
 \sigma_{r,i}^{D(3)}(Q^2, \beta, \xpom) 
  =
  \frac{ N_i^{DATA} \cdot ( 1- B)}{N^{MC}_{RAD,i}} \cdot \sigma_{r,i}^{D(3),BORN}(Q^2, \beta, \xpom)~,
\end{equation}
is measured in the kinematic domain of 
$4.5 < Q^2 < 100$ GeV$^2$, $0.008 < \beta < 1$ and $0.009 < \xpom < 0.026$. 
$N_i^{DATA}$ is the number of selected diffractive events in bin $i$, $B = 0.02$ is the global subtraction of the beam-gas background,
$N_{RAD,i}^{MC}$ is number of simulated events using MC with QED radiative corrections and $\sigma_{r,i}^{D(3),BORN}$
is a theoretical prediction of the $\rcs$ at the Born level, averaged over the respective bin volume.
\begin{figure}
\begin{center}
\includegraphics[width=0.99\columnwidth]{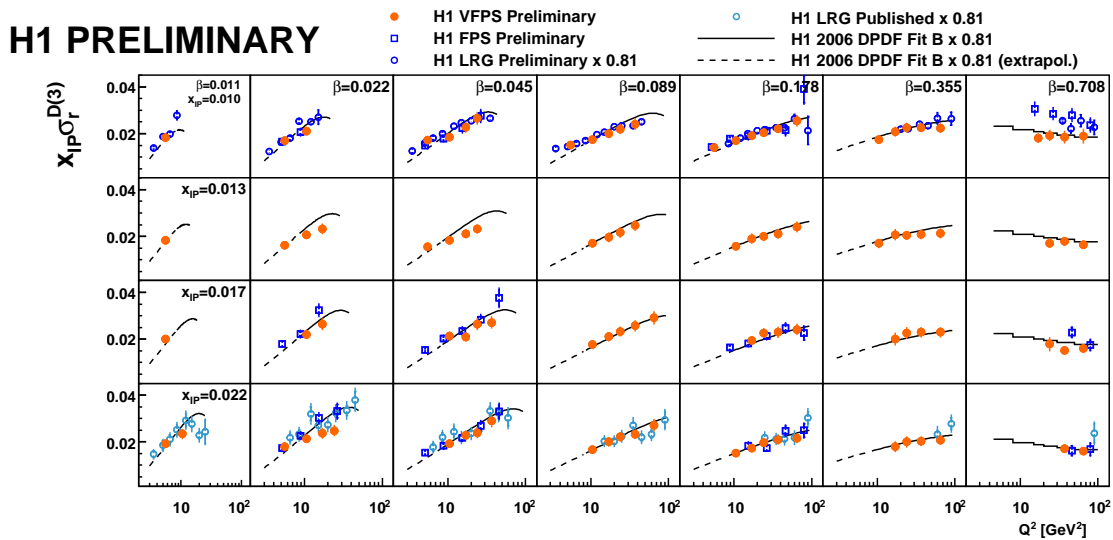}
\caption{Diffractive reduced cross section as a function of $Q^2$ in different bins of $\xpom$ and $\beta$.}
\label{Fig:RCSQ2}
\end{center}
\end{figure}
The measured $\rcs$ is shown in Fig.~\ref{Fig:RCSQ2} as a function of $Q^2$ and compared to H1 measurements
and to the prediction from H1 2006 DPDF Fit B, scaled to proton mass. At $\beta = 0.708$ bin-averaged values
of the theoretical prediction are used due to its non-trivial behaviour at large $\beta$.
A higher precision achieved in $\xpom$ reconstruction using the VFPS allowed a finer binning in $\xpom$. 
The VFPS measurement of $\rcs$ is in a remarkable agreement with both, the other H1 measurements
and H1 2006 Fit B, expressing scaling violations up to large $\beta$.
A cross-check of the reduced cross section has been made with the FPS measurement \cite{Aktas:2006hx}, where the ratio 
$\rcs($VFPS$) / \rcs($FPS$) = 0.96 \pm 0.02$ (stat) $\pm 0.11$ (syst) $\pm 0.08$ (norm) and is stable within measured points.

\end{document}